\documentclass[journal]{IEEEtran}
\usepackage{amsfonts}
\usepackage{amssymb}
\usepackage[cmex10]{amsmath}
\usepackage{graphicx}
\usepackage{subfigure}
\usepackage{verbatim}
\usepackage{cite}
\usepackage{diagbox}
\usepackage{tabularx,booktabs}
\usepackage{amsthm}
\usepackage{booktabs}
\usepackage[ruled,lined]{algorithm2e}
\usepackage{multirow}
\usepackage{setspace}
\usepackage{stfloats}
\usepackage{fancyhdr}
\usepackage{float}
\usepackage{algorithmic}
 \usepackage{enumerate}

\IEEEoverridecommandlockouts
\begin{document}

\newcommand{\be}{\begin{equation}}
\newcommand{\ee}{\end{equation}}
\newcommand{\br}{{\mbox{\boldmath{$r$}}}}
\newcommand{\bp}{{\mbox{\boldmath{$p$}}}}
\newcommand{\bpi}{\mbox{\boldmath{ $\pi $}}}
\newcommand{\bn}{{\mbox{\boldmath{$n$}}}}
\newcommand{\balfa}{{\mbox{\boldmath{$\alpha$}}}}
\newcommand{\ba}{\mbox{\boldmath{$a $}}}
\newcommand{\bta}{\mbox{\boldmath{$\beta $}}}
\newcommand{\bg}{\mbox{\boldmath{$g $}}}
\newcommand{\bPsi}{\mbox{\boldmath{$\Psi $}}}
\newcommand{\bsigma}{\mbox{\boldmath{ $\Sigma $}}}
\newcommand{\bGamma}{{\bf \Gamma }}
\newcommand{\bA}{{\bf A }}
\newcommand{\bP}{{\bf P }}
\newcommand{\bX}{{\bf X }}
\newcommand{\bI}{{\bf I }}
\newcommand{\bR}{{\bf R }}
\newcommand{\bZ}{{\bf Z }}
\newcommand{\bz}{{\bf z }}
\newcommand{\bx}{{\mathbf{x}}}
\newcommand{\bM}{{\bf M}}
\newcommand{\bU}{{\bf U}}
\newcommand{\bD}{{\bf D}}
\newcommand{\bJ}{{\bf J}}
\newcommand{\bH}{{\bf H}}
\newcommand{\bK}{{\bf K}}
\newcommand{\bm}{{\bf m}}
\newcommand{\bN}{{\bf N}}
\newcommand{\bC}{{\bf C}}
\newcommand{\bL}{{\bf L}}
\newcommand{\bF}{{\bf F}}
\newcommand{\bv}{{\bf v}}
\newcommand{\bSigma}{{\bf \Sigma}}
\newcommand{\bS}{{\bf S}}
\newcommand{\bs}{{\bf s}}
\newcommand{\bO}{{\bf O}}
\newcommand{\bQ}{{\bf Q}}
\newcommand{\btr}{{\mbox{\boldmath{$tr$}}}}
\newcommand{\bNSCM}{{\bf NSCM}}
\newcommand{\barg}{{\bf arg}}
\newcommand{\bmax}{{\bf max}}
\newcommand{\test}{\mbox{$
\begin{array}{c}
\stackrel{ \stackrel{\textstyle H_1}{\textstyle >} } { \stackrel{\textstyle <}{\textstyle H_0} }
\end{array}
$}}
\newcommand{\tabincell}[2]{\begin{tabular}{@{}#1@{}}#2\end{tabular}}
\newtheorem{Def}{Definition}
\newtheorem{Pro}{Proposition}
\newtheorem{Exa}{Example}
\newtheorem{Rem}{Remark}

\title{ Multi-Sensor Control for Multi-Target Tracking Using Cauchy-Schwarz Divergence}
\author{\IEEEauthorblockN{Meng Jiang, Wei Yi and Lingjiang~Kong }\\
\IEEEauthorblockA{School of Electronic Engineering\\
University of Electronic Science and Technology of China\\
Email:\{kussoyi, lingjiang.kong\}@gmail.com \\
}}


\maketitle
 \thispagestyle{empty}
\begin{abstract}
The paper addresses the problem of multi-sensor control for multi-target tracking via labelled random finite sets (RFS) in the sensor network systems. Based on
an information theoretic divergence measure, namely Cauchy-Schwarz (CS) divergence which admits a closed form solution for GLMB densities, we propose two novel multi-sensor control approaches in the framework of generalized Covariance Intersection (GCI). The first joint decision making (JDM) method is optimal and can achieve overall good performance, while the second independent decision making (IDM) method is suboptimal as a fast realization with smaller amount of computations. Simulation
in challenging situation is presented to verify the effectiveness of the two proposed approaches.

\end{abstract}

\section{Introduction}
Sensor network systems have received tremendous attention in last decade due to their successful applications that range from vehicular network to battlefield detection and tracking \cite{ogren2004cooperative}. In many practical situations, due to communication and computational constraints, it is required that limited amounts of sensors take right actions.
In such cases, the problem of sensor control is to find a member of the command set that can result in best measurements for filtering
purposes \cite{krishnamurthy2002algorithms}. In general, sensor control comprises two underlying components, a multi-target filtering process in conjunction with an optimal decision-making method.

Multi-target filtering has been recently investigated in a more principled way due to the point process theory or finite set statistics (FISST) based multi-target tracking methodology \cite{mahler2007statistical}. Among these random finite set (RFS) based methods, the promising generalized labeled multi-Bernoulli (GLMB) filter \cite{vo2013labeled,vo2014labeled}, or simply the Vo-Vo filter, possesses some useful analytical properties \cite{beard2015void} and is a closed form solution to the Bayes multi-target filter, can not only produce trajectories formally but also outperform the probability hypothesis density (PHD) filter \cite{vo2006gaussian}, cardinalized PHD (CPHD) filter \cite{vo2007analytic} and multi-Bernoulli (MB) filter \cite{vo2009cardinality}.

Another important component of sensor control solutions is a decision-making process, which mostly resorts to optimization of an objective function and generally falls into two categories. The first one is task-based approach, sensor control methods are designed with a direct focus on the expected performance and the objective function is formulated as a cost function, examples of such cost functions include estimated target cardinality variance \cite{hoang2014sensor,gostar2013multi}, posterior expected error of cardinality and states (PEECS) \cite{gostar2013robust,gostar2015multi} and optimal sub-pattern assignment (OSPA) distance \cite{gostar2015ospa}.
The task-based approach is useful in some situations especially where the objective function can be formulated in the form of a single criterion, but there is a challenging problem in the case of multiple competing objectives. To solve or avoid this problem, the second one is information-based approach which strives to quantify the information content of the multi-target distribution, aims at obtaining superior overall performance across multiple task objectives and the objective function is formulated as a reward function. The most common choices of reward functions are based on some information theoretic divergence measures such as Kullback¨CLeibler (KL) divergence \cite{kastella1997discrimination,aughenbaugh2008metric} and more generally the R\'{e}nyi divergence \cite{kreucher2005comparison,ristic2010sensor,ristic2011note}. However, a major limitation of utilizing KL or R\'{e}nyi divergence is their significant computational cost, and hence most of the time, one has to resort to numerical integration methods such as Monte Carlo (MC) method to derive analytically results.
An alternative information divergence measure is the Cauchy-Schwarz (CS) divergence. Using this measure, Hoang \textit{et al} provided tractable formulations
between the probability densities of two Poisson point processes \cite{hoang2015cauchy}, later, Beard \textit{et al} extended the results to two GLMB densities \cite{beard2015sensor,beard2015void} and presented an analytic expression, which opened the door to sensor control scheme with GLMB Models based on information-based approach. The CS control with GLMB models accounts for target trajectories in a principled manner, which is not possible using other tracking methods.

When the surveillance area is very large or targets move in complex movement, one sensor with limited sensing range (LSR) is not competent to the task of multi-target tracking, sensor network systems and subsequent multiple sensor control are necessary.
Inspired by the good performance achieved by sensor control with GLMB models based on CS divergence, where Beard \textit{et al} only considered single sensor,
in this paper, we address the problem of multi-sensor control for multi-target tracking using CS divergence via labelled random finite sets (RFS). To be specific, we use Vo-Vo filter to ensure local tracking performance, and Generalized Covariance Intersection (GCI) fusion \cite{mahler2000optimal,fantacci2015consensus,wang2015distributed} to maximize information content of the multi-target distribution. The key contributions of this paper are two tractable approaches of multi-sensor control, the one is optimal with a little complex calculation and the other is suboptimal as a fast realization. Simulation results verify both proposed approaches can perform well in complex situation.

\section{Background}
This section provides background material on labelled multi-target filtering, GCI fusion and Cauchy-Schwarz divergence which are necessary for the results of this paper. For further details, we refer the reader to \cite{vo2013labeled,fantacci2015consensus,wang2015distributed,beard2015void}.

\subsection{Notation }
In this paper, we adhere to the convention that single-target states are denoted by the small letters, e.g.,$x, \mathbf{x}$ while multi-target states are denoted by capital letters, e.g.,$X, \mathbf{X}$. Symbols for labeled states and their distributions/statistics (single-target or multi-target) are bolded to distinguish them from unlabeled ones, e.g., $\mathbf{x}, \mathbf{X}, \bpi$, etc. To be more specific, the labeled single target state $\bx$ is constructed by augmenting a state $x\in\mathbb{X}$ with a label $\ell\in\mathbb{L}$. Observations generated by single-target states are denoted by the small letter, e.g., $z$, and the multi-target observations are denoted by the capital letter, e.g., $Z$.
Additionally, blackboard bold letters represent spaces, e.g., the state space is represented by $\mathbb{X}$, the label space by $\mathbb{L}$, and the observation space by $\mathbb{Z}$. The collection of all finite sets of $\mathbb{X}$ is denoted by $\mathcal{F}(\mathbb{X})$.

Moreover, in order to support arbitrary arguments like sets, vectors and integers, the generalized Kronecker delta function is given by
\begin{equation}\label{delta}
  \delta_Y(X)\triangleq\left\{\begin{array}{l}
\!\!1, \,\,\,\,\mbox{if $X = Y$} \\
\!\!0, \,\,\,\,\mbox{otherwise}
\end{array}\right.
\end{equation}
and $\int\cdot\,\,\delta X$ denotes the set integral \cite{mahler2007statistical} defined by
\begin{equation}\label{set integral}
  \int\! f(X)\delta X\!=\sum_{n=0}^\infty \frac{1}{n!}\int\! f(\{x_1,\cdots,x_n\})dx_1\cdots dx_n
\end{equation}
\subsection{GLMB RFS }
An important labeled RFS is the GLMB RFS \cite{vo2013labeled}, which is a class of tractable models for on-line Bayesian inference \cite{mahler2007statistical} that
alleviates the limitations of the Poisson model. Under the standard multi-object model, the GLMB is a conjugate prior that is also closed under the Chapman-Kolmogorov equation.

Let $\mathcal{L}:\mathbb{X}\times\mathbb{L}\rightarrow\mathbb{X}$ be the projection $\mathcal{L}((x,\ell))=\ell$, and $\Delta(\bX)=\delta_{|\bX|}(|\mathcal{L}(\bX)|)$ denote the distinct label indicator.
A GLMB is an RFS on $\mathbb{X}\times\mathbb{L}$ distributed according to
\begin{equation}\label{GLMB}
\bpi(\bX)=\Delta(\bX)\sum_{c\in\mathbb{C}}w^{(c)}(\mathcal{L}(\bX))[p^{(c)}]^{\bX}
\end{equation}
where $\mathbb{C}$ is a discrete index set. The weights $w^{(c)}(L)$ and the spatial distributions $p^{(c)}$ satisfy the normalization conditions
\begin{equation*}\label{GLMB_1}
\begin{split}
\sum_{L\subseteq\mathbb{L}}\sum_{c\in\mathbb{C}}w^{(c)}(L)&=1 \\
\int p^{(c)}(x,\ell)dx&=1
\end{split}
\end{equation*}

Further, a $\delta$-GLMB RFS \cite{vo2013labeled,vo2014labeled} with state space $\mathbb{X}$ and (discrete) label space $\mathbb{L}$ is a special case of a GLMB RFS with
\begin{equation*}
\begin{split}
\mathbb{C}&=\mathcal{F}(\mathbb{L})\times\Xi  \\
w^{(c)}(L)&=w^{(I,\xi)}\delta_I(L) \\
p^{(c)}&=p^{(I,\xi)}=p^{(\xi)}
\end{split}
\end{equation*}
where $\Xi$ is a discrete space, $\xi$ are realizations of $\Xi$, and $I$ denotes a set of track labels. In target tracking applications, the discrete space $\Xi$ typically represents the history of track to measurement associations. A $\delta$-GLMB RFS is thus a special case of a GLMB RFS but with a particular structure on the index space which arises naturally in target tracking applications. The $\delta$-GLMB RFS has density
\begin{equation}\label{d-GLMB}
\bpi(\bX)=\Delta(\bX)\sum_{(I,\xi)\in\mathcal{F}(\mathbb{L})\times\Xi}w^{(I,\xi)}\delta_I(\mathcal{L}(\bX))[p^{(\xi)}]^{\bX}
\end{equation}

\subsection{Cauchy-Schwarz Divergence }
Compared with Kullback-Leibler divergence or R\'{e}nyi divergence, which are most commonly used measures of information gain, CS divergence \cite{beard2015void,beard2015sensor} has a mathematical form which is more amenable to closed form solution.

Using the relationship between probablity density and belief density, the CS divergence between two RFSs, with respective belief densities $\phi$ and $\varphi$, is given by
\begin{equation}\label{cs_rfs}
\begin{split}
  D_{CS}(\phi,\varphi)=-\ln\frac{ \int K^{|X|}\phi(X)\varphi(X)\delta X }{ \sqrt{\int K^{|X|}\phi^2(X)\delta X\int K^{|X|}\varphi^2(X)\delta X} }
\end{split}
\end{equation}
where $K$ is the unit of hyper-volume in $\mathbb{X}$.

In particular, Cauchy-Schwarz divergence has a closed form for GLMB densities, in the case where the individual target densities are Gaussian mixtures. For two GLMBs with belief densities
\begin{eqnarray}
\phi(\bX) &=& \Delta(\bX)\sum_{c\in\mathbb{C}}w_{\phi}^{(c)}(\mathcal{L}(\bX))[p_{\phi}^{(c)}]^{\bX} \\
\psi(\bX) &=& \Delta(\bX)\sum_{d\in\mathbb{C}}w_{\psi}^{(d)}(\mathcal{L}(\bX))[p_{\psi}^{(d)}]^{\bX}
\end{eqnarray}
the Cauchy-Schwarz divergence between $\phi$ and $\psi$ is given by
\begin{equation}\label{cs_glmb}
\begin{split}
D_{CS}(\phi,\psi)=-\ln\frac{ \zeta(\phi,\psi) }{ \sqrt{\zeta(\phi,\phi)\zeta(\psi,\psi)} }
\end{split}
\end{equation}
where
\begin{equation}\
\begin{split}
\zeta(\phi,\psi)= & \sum_{L\subseteq\mathbb{L}} \sum_{c\subseteq\mathbb{C}} \sum_{d\subseteq\mathbb{D}} w_\phi^{(c)}(L)w_\psi^{(d)}(L) \\
&\times[K\int p_\phi^{(c)}(x,\cdot) p_\psi^{(d)}(x,\cdot)dx]^L
\end{split}
\end{equation}

Closed form of the analytical expression using CS divergence combines GLMB densities and information theoretic divergence measures hence leads to a more efficient implementation of sensor control.

\subsection{Distributed Fusion }
In the context of sensor network systems with LRS, where each sensor has a finite field of view (FoV), distributed fusion is necessary to make the best use of local distribution information in order to solve the shadowing effect.
The GCI was proposed by Mahler \cite{mahler2000optimal} specifically to extend FISST to sensor network systems, which is capable to fuse both Gaussian and non-Gaussian formed multi-target distributions from different sensor with completely unknown correlation.

Based on GCI, with the assumption that all the sensor nodes share the same label space for the birth process, Fantacci \textit{et al} proposed the GCI fusion with labeled set filters by use the consistent label. The results include consensus marginalized $\delta$-GLMB (CM$\delta$-GLMB) and consensus LMB (CLMB) tracking filter \cite{fantacci2015consensus}.

\subsubsection{CM$\delta$-GLMB }

Suppose that each sensor $i=1,\ldots,N$ is provided with an M$\delta$-GLMB density $\bpi^i$ of the form
\begin{equation}\label{MGLMB}
\bpi^i=\Delta(\bX)\sum_{L\in\mathcal{F}(\mathbb{L})}\delta_L(\mathcal{L}(\bX))w_i^{(L)} [p_i^{(L)}]^{\bX}
\end{equation}
where $N$ is the total sensor number and fusion weight $\omega^i\in(0,1)$, $\sum_{i=1}^N\omega^i=1$, then the fused distribution is given as follows:
\begin{equation}\label{CMGLMB}
\bpi^s=\Delta(\bX)\sum_{L\in\mathcal{F}(\mathbb{L})}\delta_L(\mathcal{L}(\bX))w_s^{(L)} [p_s^{(L)}]^{\bX}
\end{equation}
where
\begin{eqnarray*}
w_s^{(L)} &=& \frac{ \displaystyle{\prod_{i=1}^N}\left(w_i^{(L)}\right)^{\omega^i}
\left[ \int\displaystyle{\prod_{i=1}^N}\left(p_i^{(L)}(x,\cdot)\right)^{\omega^i}dx \right]^L }
{ \displaystyle{\sum_{F\in\mathbb{L}}\prod_{i=1}^N} \left(w_i^{(F)}\right)^{\omega^i} \left[ \int\displaystyle{\prod_{i=1}^N}\left(p_i^{(F)}(x,\cdot)\right)^{\omega^i}dx \right]^F }   \\
p_s^{(L)} &=& \frac{ \displaystyle{\prod_{i=1}^N}\left(p_i^{(L)}\right)^{\omega^i}  }
{ \displaystyle{\int\prod_{i=1}^N}\left(p_i^{(L)}\right)^{\omega^i}dx }
\end{eqnarray*}

\subsubsection{CLMB }
Suppose that each sensor $i=1,\ldots,N$ is provided with a LMB density $\bpi^i$ of the form $\{( r_i^{(\ell)},p_i^{(\ell)} )\}_{\ell\in\mathbb{L}}$, where $N$ is the total sensor number and fusion weight $\omega^i\in(0,1)$, $\sum_{i=1}^N\omega^i=1$, then the fused distribution is of the form
\begin{equation}\label{CLMB}
\bpi^s=\{( r_s^{(\ell)},p_s^{(\ell)} )\}_{\ell\in\mathbb{L}}
\end{equation}
where
\begin{eqnarray*}
r_s^{(\ell)} &=& \frac{ \displaystyle{\int\prod_{i=1}^N}\left(r_i^{(\ell)}p_i^{(\ell)}(x)\right)^{\omega^i}dx }
{ \displaystyle{\prod_{i=1}^N}\left(1-r_i^{(\ell)}\right)^{\omega^i}+\displaystyle{\int\prod_{i=1}^N}\left(r_i^{(\ell)}p_i^{(\ell)}(x)\right)^{\omega^i}dx }\\
p_s^{(\ell)} &=& \frac{ \displaystyle{\prod_{i=1}^N}\left(p_i^{(\ell)}\right)^{\omega^i}  }
{ \displaystyle{\int\prod_{i=1}^N}\left(p_i^{(\ell)}\right)^{\omega^i}dx }
\end{eqnarray*}

Consensus algorithms can fuse in a fully distributed and scalable way the information collected from the multiple heterogeneous and geographically dispersed sensors, and therefore have a significant impact on the estimation performance of the tracking system.

\section{Multi-Sensor Control Using CS divergence}
In most target tracking scenarios, the sensor may perform various actions that can maximize the tracking observability, and can therefore influence the estimation performance of the tracking system. Typically, such actions may include changing the position, altering the sensor operating parameters, orientation or motion of the sensor platform and so on, which in turn affects the sensor's ability to detect and track targets.

In the context of sensor network systems, where there are more than one sensor waiting to be deployed, the allowable control actions may increase exponentially and hence
the control of multi-sensor is a high-dimensional optimization problem. Therefore, making control decisions by manual intervention or some deterministic control policy which provides no guarantee of optimality, is not a good choice.
Compared with single sensor control, there are some challenging problems in multi-sensor control such as aforementioned high-dimensional optimization problem and information fusion problem induced by the measurement collected from the multiple sensors.
In this section, we seek tractable solution for multi-sensor control for multi-target tracking with GLMB models.

\subsection{Problem Formulation}
In sensor network systems, one or more sensors are the direct outputs of the decision-making component of the control solution, as such, the focus has traditionally been placed on improving the decision-making component. However, the multi-target tracking component also plays a significant role in the overall performance of the scheme in terms of accuracy and robustness.

Inspired by the versatile GLMB model which offers good trade-offs between tractability and fidelity, in filtering stage, we use the Vo-Vo filter \cite{vo2013labeled,vo2014labeled} as local sensor and GCI fusion to fuse the information collected from the multiple sensors in order to achieve overall superior performance, the procedure is described as follows:

1) At time step $k$, with measurement $Z^{i}_k=\{z^i_{1,k},z^i_{2,k},\ldots,z^i_{m,k}\}$ where the subscript $k$ denotes current time and superscript $i$ denotes sequence number of sensors,
each sensor node $i=1,\ldots, N$ locally performs prediction and update using Vo-Vo filter, the details can be found in \cite{vo2014labeled}.

2) Implement the GCI fusion with local posterior distribution $\bpi_k^i$ to derive the fused distribution $\bpi_k^s$, the superscript $s$ denotes fused distribution.
Note that one needs to convert $\delta$-GLMB posterior distribution to M$\delta$-GLMB$\backslash$LMB distribution for consensus fusion method using (\ref{CMGLMB}) or (\ref{CLMB}).

3) After fusion, an estimate of the object set $\hat{X}_{k|k}$ is obtained from the cardinality probability mass function and the location PDFs using MAP technique.

A pseudo-code of filtering stage is given in Algorithm 1.

\begin{algorithm}[htb]
\DontPrintSemicolon
\caption{\label{alg1} Filtering Procedure\;}
\small\underline{\sc \textbf{Input}}: $\bpi_{k-1}^i\{\bX|Z^i_{1:k-1}\}, Z^{i}_k$ \;
\small\underline{\sc \textbf{Output}}: $\bpi_k^i\{\bX|Z^i_{1:k}\}$,$\bpi_k^s$\;
\For{ $i=1:N$ }{
 local prediction\;
 local update $\rightarrow\bpi_k^i\{\bX|Z^i_{1:k}\} $\;
}
GCI($\bpi_k^i\{\bX|Z^i_{1:k}\}$)$\rightarrow\bpi_k^s $\;
MAP($\bpi_k^s$ )$\rightarrow\hat{X}_{k|k}$\;
\end{algorithm}

In control strategy, we adhere to the convention that formulating the sensor control problem as a Partially Observed Markov Decision Process (POMDP) using FISST \cite{mahler1998global} and defining the following notation: $\bpi^i_k(\cdot|Z^i_{1:k})$ is the posterior density for sensor $i$ at time $k$, $\mathbb{C}_i$ is the control action space for sensor $i$ and hence the $N$ multiple sensor control action space $\mathbb{C}=\mathbb{C}_1\times\cdots\times \mathbb{C}_N$, $H$ is the length of control horizon, the $\bpi^i_{k+H}(\cdot|Z^i_{1:k})$ is predicted density at time $k+H$ based on known measurements from time 1 to time $k$, $\tilde{Z}^i_{k+1:k+H}(c_{1},\ldots,c_{N})$ is the collection of measurements for sensor $i$ that would be observed from times $k+1$ up to $k+H$ with executed control action $(c_{1},\ldots,c_{N})\in \mathbb{C}$ at time $k$, note that $c_i\in\mathbb{C}_i$ is a vector composed of all possible actions what a sensor can take, such as changing direction of movement, velocity, power and so on.

We use CS divergence as reward function at the control horizon which is measured between the predicted and posterior multi-target density:

\begin{equation}\label{reward_pu}
\begin{split}
R(c_{1},\ldots,&c_{N})=D_{CS}( \bpi_{\rm prediction},\bpi_{\rm update} )
\end{split}
\end{equation}
then the optimal control action is decided by maximising the expected value of the reward function $R(c_{1},\ldots,c_{N})$ over the allowable actions space $\mathbb{C}$:
\begin{equation}\label{optimal}
(\hat{c}_{1},\ldots,\hat{c}_{N})={\rm arg} \max \limits_{(c_{1},\ldots,c_{N})\in \mathbb{C}} {\rm EAP}(R(c_{1},\ldots,c_{N}))
\end{equation}

Note that the above expected reward is not available to analytic solutions, so we resort to Monte Carlo integration,
\begin{equation}\label{optimal_apro}
{\rm EAP}(R(c_{1},\ldots,c_{N}))\approx \frac{1}{M}\sum_{j=1}^M R^{(j)}(c_{1},\ldots,c_{N})
\end{equation}
where $M$ denotes the number of samples. Also for this reason, we prefer CS divergence which provides a closed-form solution with GLMB models to calculate $R^{(j)}(c_{1},\ldots,c_{N})$, can alleviate the side effect induced by the Monte Carlo technique (\ref{optimal_apro}).

In what following we detail the design of predicted distribution and posterior distribution in (\ref{reward_pu}) and present two multi-sensor control approaches.

\subsection{Multi-Sensor Control Strategy }

\vspace{3mm}
\noindent{\small\underline{\sc \textbf{Joint Decision Making Algorithm}}}

In order to make the best use of sensor network and overall collected information, we propose an optimal multi-sensor control approach, referred to joint decision making (JDM) algorithm. In this method, the filtering stage is performed as described in Algorithm 1, the fused density $\bpi_k^s$ will be used for multi-target samples in order to solve the shadowing effect of single sensor with LSR and to compute the predicted density at the end of the control horizon. The specific procedure are as follows:

1) Multi-target Samples: At desicion time step $k$, draw a set $\Psi_S$ of $M$ multi-target samples from fused distribution $\bpi_k^s$, it is mainly designed for deriving numerical analytical resolutions of CS reward function.

2) Pseudo-Prediction: Compute the predicted density at the end of the control horizon $\bar{\bpi}_{k+H}^s$, which will be later used as one term
of computing CS divergence, by carrying out repeated prediction steps of Vo-Vo filter, without traget birth or death, for this reason, we use the term ``pseudo-prediction".

3) Generate predicted ideal measurement (PIMS): For each sensor $i=1,\ldots,N$ and each multi-target sample $\bX^{(j)}\in \Psi_S$, generating PIMS $\tilde{Z}_{k+1:k+H}^{i}(c_i,\bX^{(j)})$ with current control action $c_i\in \mathbb{C}_i$ based on initial predicted trajectory in sample $\bX^{(j)}$,
more detials in \cite{beard2015sensor,mahler2005multitarget}.

4) Run Vo-Vo Filter Recursion: Run each Vo-Vo filter with initial local posterior distribution $\bpi_k^i\{\bX|Z^i_{1:k}\}$ using PIMS $\tilde{Z}_{k+1:k+H}^{i}(c_i,\bX^{(j)})$ to get the pseudo updated distribution $\bpi_{k+H}^i\{\bX|Z^i_{1:k},\tilde{Z}_{k+1:k+H}^{i}(c_i,\bX^{(j)})\}$,
we will use the term ``filter'' to denote Vo-Vo filter recursion \cite{vo2014labeled}.

5) GCI Fusion: For multi-sensor, for each possible control action combination $(c_{1},\ldots,c_{N})\in \mathbb{C}$, perform the GCI fusion with pseudo updated distribution $\bpi_{k+H}^i\{\bX|Z^i_{1:k},\tilde{Z}_{k+1:k+H}^{i}(c_i,\bX^{(j)})\}$ to get the fused pseudo updated distribution $\bpi_{k+H}^s(c_{1},\ldots,c_{N},\bX^{(j)})$, it will be later used as another term of computing CS divergence.

6) Compute Each Reward: Compute CS reward function for each control action combination and each sample using (\ref{cs_glmb}),
\begin{equation}\label{reward-alg1}
R^{(j)}(c_{1},\ldots,c_{N})= D_{CS}(\bar{\bpi}_{k+H}^s,\bpi_{k+H}^s(c_{1},\ldots,c_{N},\bX^{(j)}))
\end{equation}
after the computation of (\ref{reward-alg1}) for all samples in set $\Psi_S$, we then compute the expected value of the reward function
\begin{equation}\label{reward-alg1e}
\begin{split}
R(c_{1},\ldots,c_{N})&= {\rm EAP} (R^{(j)}(c_{1},\ldots,c_{N})) \\
&\approx\frac{1}{M}\sum_{j=1}^M R^{(j)}(c_{1},\ldots,c_{N})
\end{split}
\end{equation}

7) Joint Decision Making: Maximize the expected value of the reward function $R(c_{1},\ldots,c_{N})$ over the allowable action space $\mathbb{C}$ using (\ref{optimal}).

A pseudo-code of above control stage is shown in Algorithm 2.

\begin{algorithm}[htb]
\DontPrintSemicolon
\caption{\label{alg3} JDM Procedure\;}
\small\underline{\sc {Input}}: $\bpi_k^s$,$\bpi_k^i\{\bX|Z^i_{1:k}\}$,$\mathbb{C}$ \;
\small\underline{\sc {Output}}: $(\hat{c}_{1},\ldots,\hat{c}_{N})$\;

\textbf{Multi-target Samples}: \;
$\bpi_k^s\rightarrow \Psi_S=\{\bX^{(1)},\ldots\bX^{(M)}\}$ \;
\textbf{Pseudo-Prediction}: \;
\For{ $iter=k+1:k+H$ }{
$\bpi_k^s\rightarrow\bar{\bpi}_{k+H}^s$\;
}
\For{ $i=1:N$ }{
\For{ each $c_{i}\in \mathbb{C}_i$ }{
\For{ each $\bX^{(j)}\in \Psi_S$ }{
\textbf{Generate PIMS}: \;
$\bX^{(j)}\rightarrow\tilde{Z}_{k+1:k+H}^{i}(c_i,\bX^{(j)})$\;
\textbf{Run Vo-Vo Filter Recursion}: \;
filter($\bpi_k^i\{\bX|Z^i_{1:k}\},\tilde{Z}_{k+1:k+H}^{i}(c_i,\bX^{(j)}))$
$\rightarrow\bpi_{k+H}^i\{\bX|Z^i_{1:k},\tilde{Z}_{k+1:k+H}^{i}(c_i,\bX^{(j)})\}$\;
}
}
}
\textbf{GCI Fusion}: \;
\For{ each $(c_{1},\ldots,c_{N})\in \mathbb{C}$ }{
\For{ each $\bX^{(j)}\in \Psi_S$ }{
GCI($ \bpi_{k+H}^1\{\bX|Z^1_{1:k},\tilde{Z}_{k+1:k+H}^1(c_{1},\bX^{(j)})\},\ldots,$
$\bpi_{k+H}^N\{\bX|Z^N_{1:k},\tilde{Z}_{k+1:k+H}^N(c_{N},\bX^{(j)})\} $)
$\rightarrow \bpi_{k+H}^s(c_{1},\ldots,c_{N},\bX^{(j)})$ \;
\textbf{Compute Each Reward}: \;
$D_{CS}$($\bar{\bpi}_{k+H}^s,\bpi_{k+H}^s(c_{1},\ldots,c_{N},\bX^{(j)})$)
$\rightarrow R^{(j)}(c_{1},\ldots,c_{N})$
}
EAP($R^{(j)}(c_{1},\ldots,c_{N})$) $\rightarrow R(c_{1},\ldots,c_{N})$
}
\textbf{Joint Decision Making}: \;
${\rm arg} \max \limits_{(c_{1},\ldots,c_{N})\in\mathbb{C}} (R(c_{1},\ldots,c_{N})) \rightarrow (\hat{c}_{1},\ldots,\hat{c}_{N})$

\end{algorithm}

Note that in the JDM algorithm, GCI fusion has been uesd both in filtering stage and CS control stage, aims at maximizing observation information content and overall CS divergence, to ensure multiple sensors move in direction where the overall performance is satisfying.

Moreover, in order to reduce the computation burden of the JDM algorithm, which is mainly induced by allowable control action combination with computation complexity $O(|\mathbb{C}_1|\times\cdots\times |\mathbb{C}_N|)$, one can resort to importance sampling technique, more details in \cite{morelande2009joint}.

\vspace{3mm}
\noindent{\small\underline{\sc \textbf{Independent Decision Making Algorithm }}}

We also propose another suboptimal multi-sensor control approach, referred to independent decision making (IDM) algorithm.
In this method, the filtering stage is same but the control stage is simplified as a fast implementation. In particular, the GCI fusion is only performed in filtering stage and each sensor makes control decision independently in control stage, which enables parallel execution of the control step, and therefore the
computation complexity of allowable control action is reduced to $O(|\mathbb{C}_1|+\cdots+|\mathbb{C}_N|)$. A pseudo-code of IDM algorithm is shown in Algorithm 3.

\begin{algorithm}[htb]
\DontPrintSemicolon
\caption{\label{alg2} IDM Procedure\;}
\small\underline{\sc {Input}}: $\bpi_k^s$,$\bpi_k^i\{\bX|Z^i_{1:k}\}$,$\mathbb{C}$ \;
\small\underline{\sc {Output}}: $(\hat{c}_{1},\ldots,\hat{c}_{N})$\;

\textbf{Multi-target Samples}: \;
$\bpi_k^s\rightarrow \Psi_S=\{\bX^{(1)},\ldots\bX^{(M)}\}$ \;
\textbf{Pseudo-Prediction}: \;
\For{ $iter=k+1:k+H$ }{
$\bpi_k^s\rightarrow\bar{\bpi}_{k+H}^s$\;
}
\For{ $i=1:N$ }{
\For{ each $c_i\in \mathbb{C}_i$ }{
\For{ each $\bX^{(j)}\in \Psi_S$ }{
\textbf{Generate PIMS}: \;
$\bX^{(j)}\rightarrow\tilde{Z}_{k+1:k+H}^{i}(c_i,\bX^{(j)})$\;
\textbf{Run Vo-Vo Filter Recursion}: \;
filter($\bpi_k^i\{\bX|Z_{1:k}\},\tilde{Z}_{k+1:k+H}^{i}(c_i,\bX^{(j)}))$
$\rightarrow\bpi_{k+H}^i\{\bX|Z_{1:k},\tilde{Z}_{k+1:k+H}^{i}(c_i,\bX^{(j)})\}$\;
\textbf{Compute Each Reward}: \;
$D_{CS}$($\bar{\bpi}_{k+H}^s,\bpi_{k+H}^i\{\bX|Z_{1:k},\tilde{Z}_{k+1:k+H}^{i}(c_i,\bX^{(j)})\}$)
$\rightarrow R_{i}^{(j)}(c_i)$
}
EAP($R_{i}^{(j)}(c_i))$ $\rightarrow R_{i}(c_i)$
}
\textbf{Decision Making on Each Sensor}: \;
${\rm arg} \max \limits_{c_i \in \mathbb{C}_i} (R_{i}(c_i)) \rightarrow \hat{c}_i$
}
\end{algorithm}

Note that the fused distribution $\bpi_k^s$ is used in multi-target samples and pseudo-prediction, which can ensure observability in control stage so that avoid making myopic decisions.

\begin{figure}[!h]
\begin{center}
\includegraphics[width=0.85\columnwidth,draft=false]{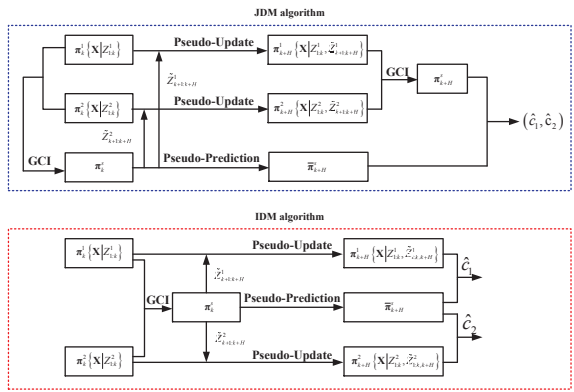}
\end{center}
\caption{A comparison between JDM algorithm and IDM algorithm with two sensors.}
\label{fig: control}
\end{figure}

A comparison between JDM algorithm and IDM algorithm with two sensors is illustrated in Fig. \ref{fig: control}.

\section{Simulation Results and Discussion}
In this section, the two proposed multi-sensor control approaches are applied to the problem of multi-target tracking with two sensors with LSR. With both methods, local filters are Vo-Vo filters, the fusion method is chosen as CM$\delta$-GLMB and fusion weight of each sensor $\omega_1,\omega_2$ are both chosen as 0.5.

The kinematic target state is a vector of planar position and velocity $x_k=[t_{x,k} \,\, \dot{t}_{x,k} \,\, t_{y,k} \,\, \dot{t}_{y,k}]^T$ and the single-target state space model is linear Gaussian according to transition density $f_{k|k-1}(x_k|x_{k-1})=\mathcal{N}(x_k,F_kx_{k-1},Q_k)$ with parameters
\begin{equation*}
F_k=\left[
  \begin{array}{ccc}
    I_2 & \Delta I_2 \\
    0_2 & I_2 \\
  \end{array}
\right]
,
Q_k=\sigma_v^2\left[
  \begin{array}{ccc}
    \frac{\Delta^4}{4}I_2 & \frac{\Delta^3}{2}I_2 \\
    \frac{\Delta^3}{2}I_2 & \Delta^2I_2 \\
  \end{array}
\right]
\end{equation*}
where $I_n$ and $0_n$ denote the $n\times n$ identity and zero matrices respectively, $\Delta=1s$ is the sampling period, $\sigma_v=5m/s^2$ is the standard deviations of the process noise.

In the context of multi-sensor control, we consider the following sensor models that the measurement as well as the detection probability is a function of distance between target and sensor states. The sensor measurements are noisy vectors of polar position of the form
\begin{equation*}\label{zk}
z_k=\left[
  \begin{array}{c}
    \arctan(\frac{t_{y,k}-s_{y,k}}{t_{x,k}-s_{x,k}}) \\
    \sqrt{(t_{x,k}-s_{x,k})^2+(t_{y,k}-s_{y,k})^2} \\
  \end{array}
\right] + w_k(x_k,u_k) \\
\end{equation*}
where $u_k=[s_{x,k} \,\, s_{y,k}]$ denotes sensor position.
$w_k(x_k,u_k)\sim\mathcal{N}(\cdot;0,R_k)$ is the measurement noise with covariance $R_k= {\rm diag}(\sigma_\theta^2,\sigma_r^2)$ in which the scales of range and bearing noise are $\sigma_r=\sigma_0+\eta_r\|x_k-u_k\|^2$ and $\sigma_\theta=\theta_0+\eta_\theta\|x_k-u_k\|$, the parameters $\sigma_0=10m$, $\eta_r=5\times10^{-5}m^{-1}$, $\theta_0=\pi/180 rad$ and $\eta_\theta=5\times10^{-6}m^{-1}$.
The probability of target detection in each sensor is independent and of the form
\begin{equation*}\label{pd}
P_D(x_k,u_k)=\frac{ \mathcal{N}(\|x_k-u_k\|;0,\sigma_D) }
{ \mathcal{N}(0;0,\sigma_D) }
\end{equation*}
where $\sigma_D=10000m$ controls the rate at which the detection probability drops off as the range increases.
Moreover, the survival probability is $P_{S,k}=0.98$, the number of clutter reports in each scan is Poisson distributed with $\lambda_c=25$. Each clutter report is sampled uniformly over the whole surveillance region.

\begin{figure}[!h]
\begin{center}
\includegraphics[width=0.85\columnwidth,draft=false]{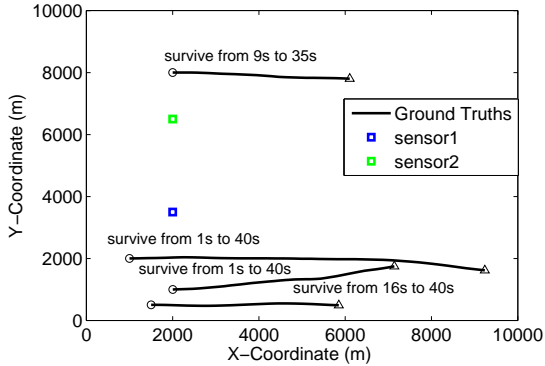}
\end{center}
\caption{Target trajectories considered in the simulation experiment. The start/end point for each trajectory is denoted, respectively, by $\circ|\bigtriangleup$.
The $\square$ indicates initial sensor position.}
\label{fig: truth}
\end{figure}

The sensor platform moves with constant velocity but takes course changes at pre-specified decision time. The allowable control actions for each sensor is $\mathbb{C}_i=[ -180^\circ, -150^\circ, \ldots, 0^\circ, \ldots, 150^\circ, 180^\circ] $, the number of samples used to compute the expected reward is $M=40$, the idealised measurements are generated over a horizon length of $H=5$, with sampling period $T=2$s.
The test scenario consists of 4 targets, the sensors keep still during first 10s and make first decision at 10s so the second decision at 20s, third decision at 30s, then remain on that course until the end of the scenario at time 40s. The region and tracks are shown in Fig. \ref{fig: truth}.

\begin{figure}[h]
\begin{minipage}[b]{0.48\linewidth}
  \centering
  \centerline{\includegraphics[width=4.8cm]{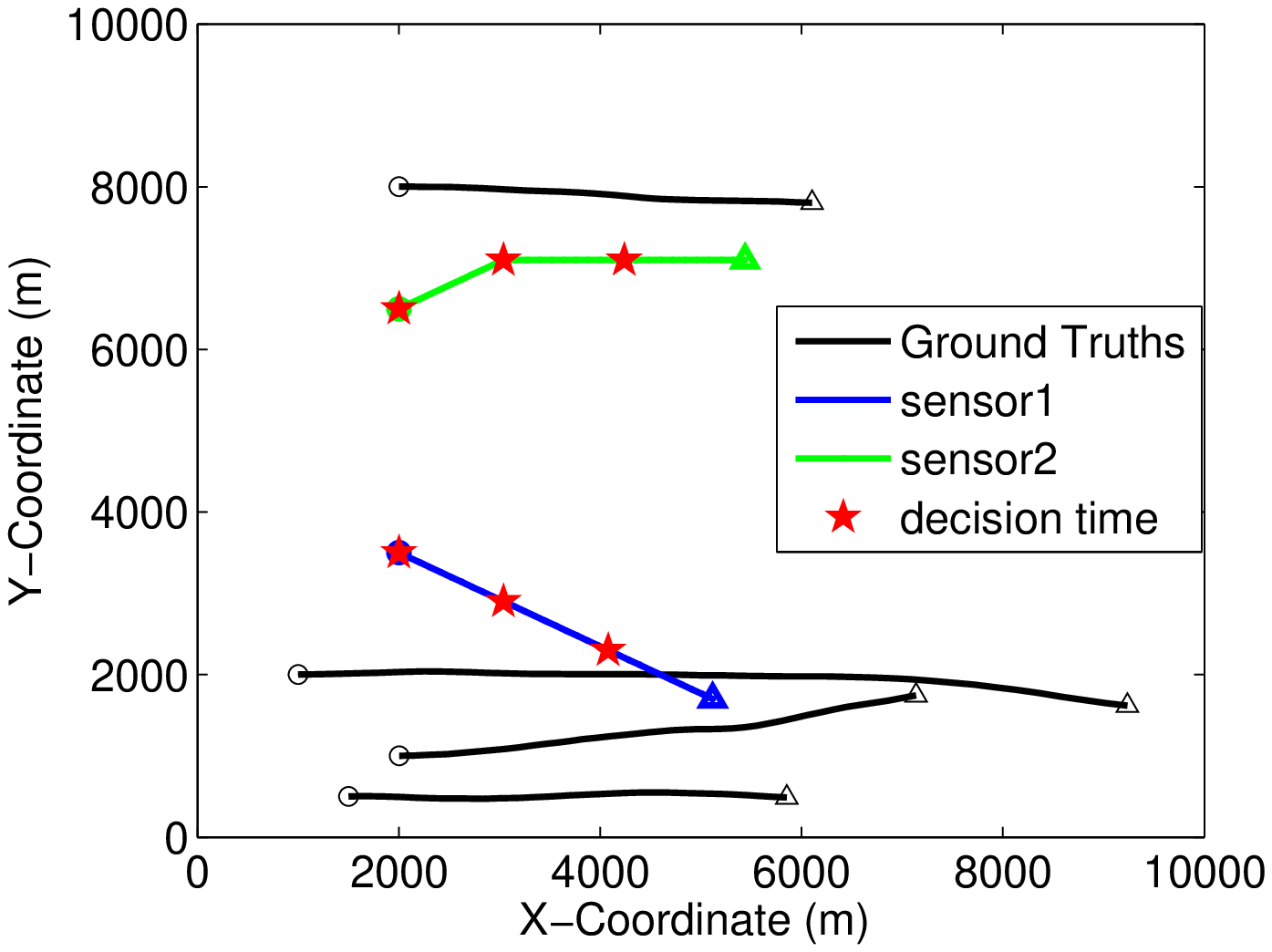}}
  \centerline{(a)}\medskip
\end{minipage}
\hfill
\begin{minipage}[b]{0.48\linewidth}
  \centering
  \centerline{\includegraphics[width=4.8cm]{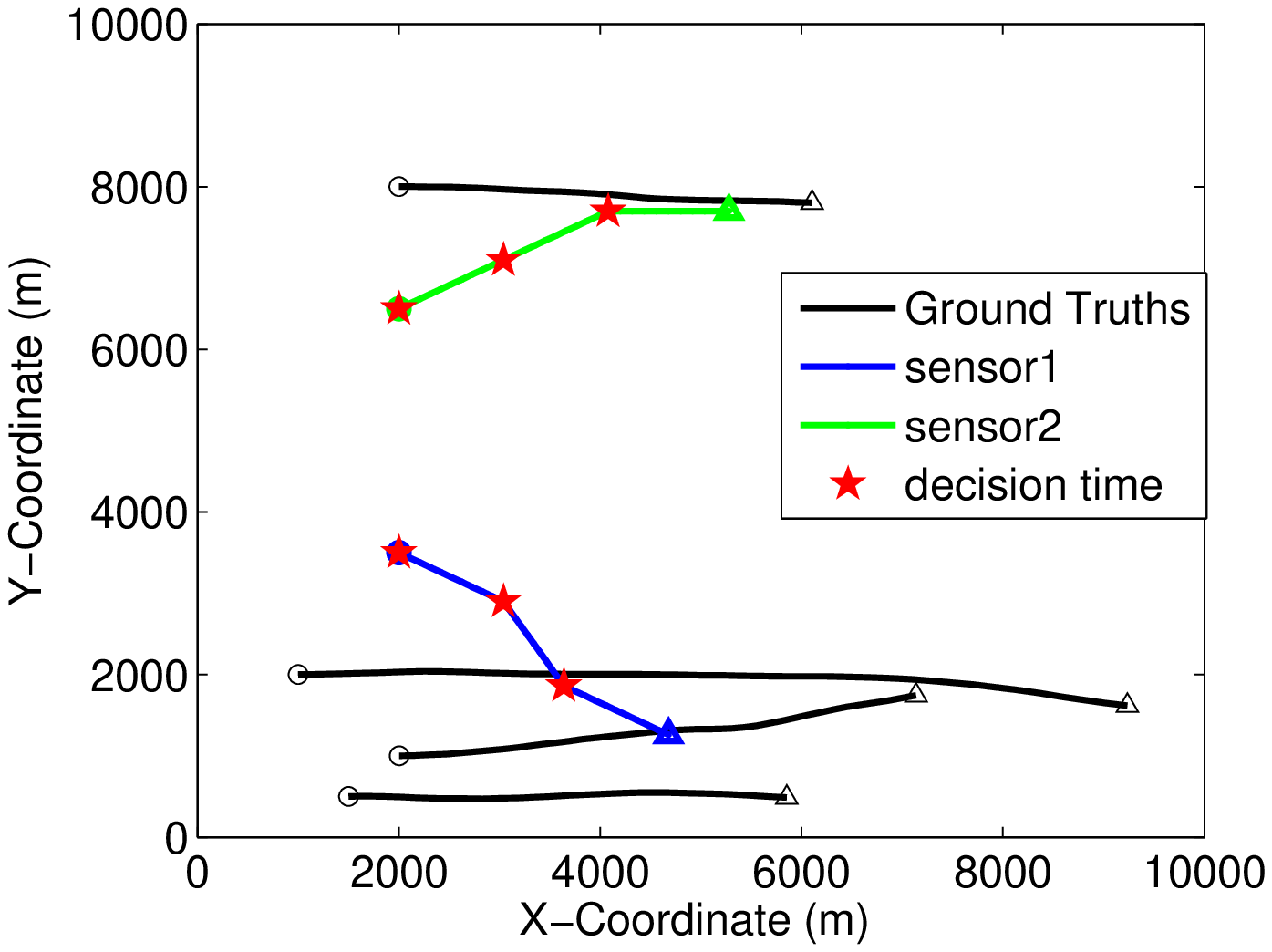}}
  \centerline{(b) }\medskip
\end{minipage}
\caption{(a) Track output from a typical run based on IDM algorithm. (b) Track output from a typical run based on JDM algorithm.}
\label{fig: c1c2truth}
\end{figure}

\begin{figure}[h]
\begin{minipage}[b]{0.48\linewidth}
  \centering
  \centerline{\includegraphics[width=4.8cm]{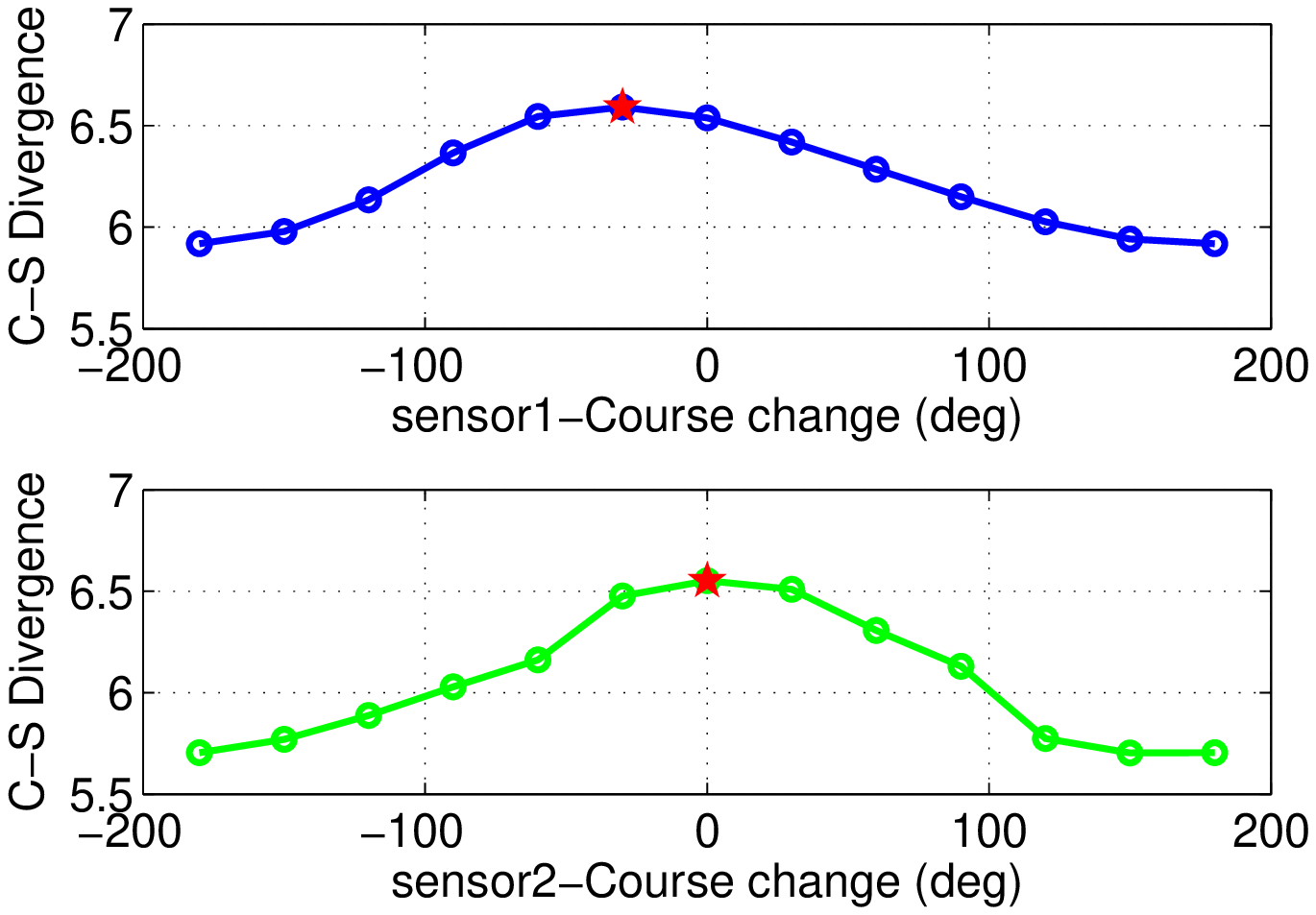}}
  \centerline{(a)}\medskip
\end{minipage}
\hfill
\begin{minipage}[b]{0.48\linewidth}
  \centering
  \centerline{\includegraphics[width=4.8cm]{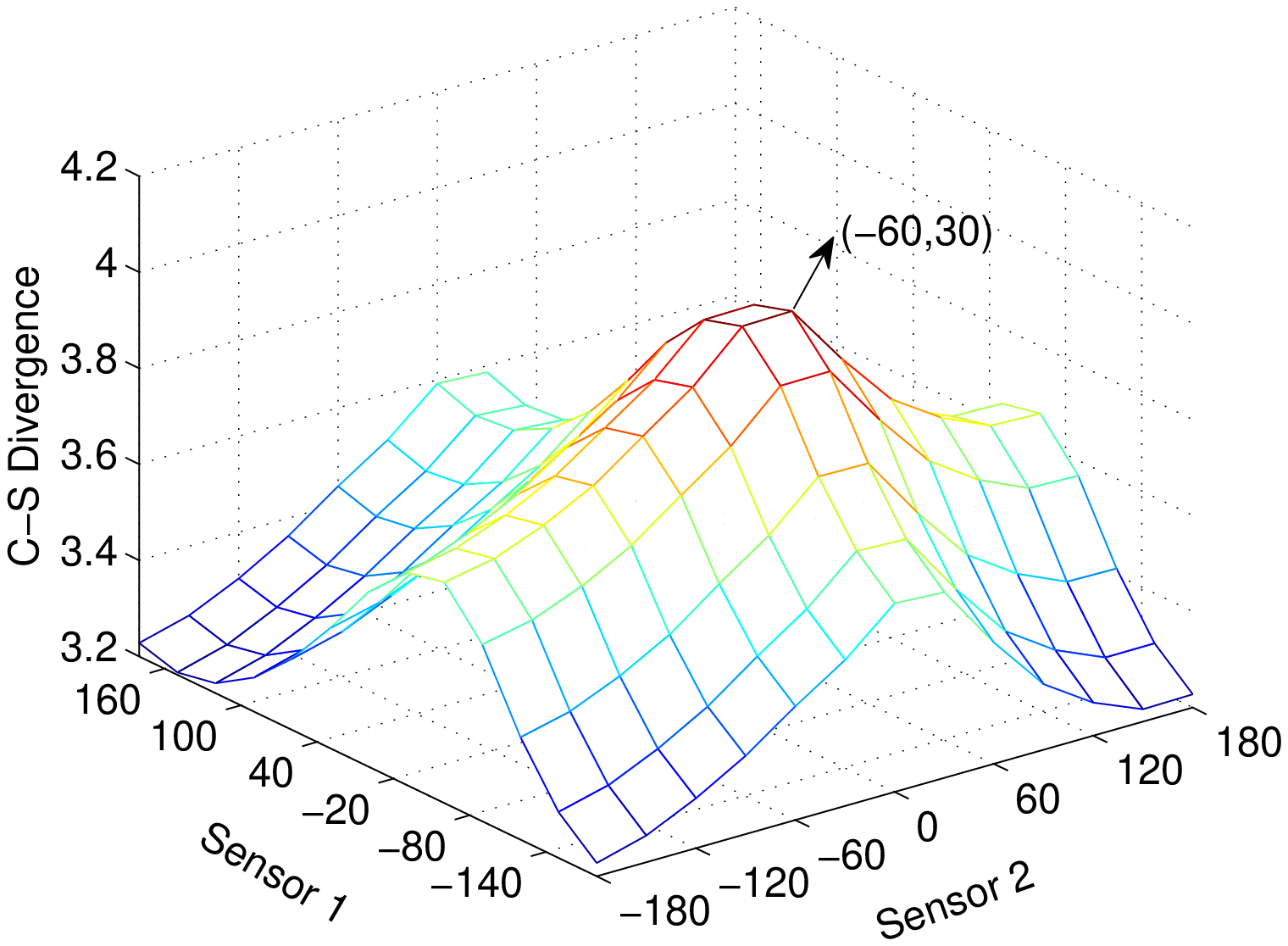}}
  \centerline{(b) }\medskip
\end{minipage}
\caption{(a) Reward curve at the time of the second decision (20s) based on IDM algorithm. (b) Reward curve at the time of the second decision (20s) based on JDM algorithm.}
\label{fig: c1c2cs}
\end{figure}

\begin{figure}[!h]
\begin{center}
\includegraphics[width=0.85\columnwidth,draft=false]{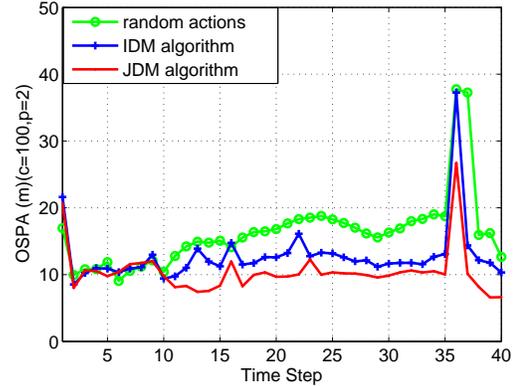}
\end{center}
\caption{Comparison of OSPA errors returned by randomised control action, IDM algorithm and JDM algorithm. The plotted results are the average of 100 Monte Carlo run.}
\label{fig: ospa}
\end{figure}

Fig. \ref{fig: c1c2truth} (a) and (b) show a single run to exhibit the typical control behaviour based on IDM algorithm and JDM algorithm, respectively. As it can be seen, both control methods can make proper decisions that sensors move close to the targets. To be more specific, we denote the control action chosen by sensor 1 and sensor 2 by a vector $(\theta_1,\theta_2)$, at the first decision time 10s, two control methods make same decision $(-30^\circ,30^\circ)$, at the second decision time 20s, the IDM algorithm takes $(-30^\circ,0^\circ)$ while the JDM algorithm takes $(-60^\circ,30^\circ)$.
Fig. \ref{fig: c1c2cs} (a) and (b) show the CS divergence at the second decision (20s) of IDM algorithm and JDM algorithm, respectively.
These results mean that compared with the IDM algorithm, each sensor controlled by JDM algorithm is not greedy to observe all targets, but rather a view of the whole picture to make the amount of information content of fused density larger.
Fig. \ref{fig: ospa} shows the comparison of OSPA errors averaged over 100 Monte Carlo runs among randomised control action, IDM algorithm and JDM algorithm. As it is shown, both control methods can achieve better performance than randomised control strategy and the JDM algorithm is preferable. Moreover, when the situation is more complex such as much more targets or sensors, the performance difference between JDM algorithm and IDM algorithm will increase and the randomised control strategy may collapse.

\section{Conclusion}
In this paper, we address the problem of multi-sensor control for multi-target tracking via labelled random finite sets (RFS) in the sensor network systems. With the GCI fusion, two novel multi-sensor control approaches using CS divergence are presented, referred to JDM and IDM algorithm, respectively. Simulation results verify both the control approaches perform well in multi-target tracking, the IDM method has smaller amount of computations while the JDM method makes decision from holistic point of view, and hence achieve better performance.

\section*{Acknowledgment}
This work was supported by the National Natural Science Foundation of China under Grants 61301266, the Chinese Postdoctoral Science Foundation under Grant 2014M550465.

\bibliographystyle{ieeetr}
\bibliography{DFCS}

\end{document}